\documentclass[pra,showpacs,twocolumn]{revtex4}

\usepackage{amsmath,amssymb,amsfonts,subfigure,dsfont}
\usepackage[english]{babel}
\usepackage[dvips]{graphicx}

\def\bra#1{\mathinner{\langle{#1}|}}
\def\ket#1{\mathinner{|{#1}\rangle}}
\def\braket#1{\mathinner{\langle{#1}\rangle}}

\let\protect\relax
{\catcode`\|=\active
  \xdef\Braket{\protect\expandafter\noexpand\csname Braket \endcsname}
  \expandafter\gdef\csname Braket \endcsname#1{\begingroup
     \ifx\SavedDoubleVert\relax
       \let\SavedDoubleVert\|\let\|\BraDoubleVert
     \fi
     \mathcode`\|32768\let|\BraVert
     \left\langle{#1}\right\rangle\endgroup}
}
\def\BraVert{\@ifnextchar|{\|\@gobble}
     {\egroup\,\mid@vertical\,\bgroup}}
\def\BraDoubleVert{\egroup\,\mid@dblvertical\,\bgroup}
\let\SavedDoubleVert\relax

{\catcode`\|=\active
  \xdef\set{\protect\expandafter\noexpand\csname set \endcsname}
  \expandafter\gdef\csname set \endcsname#1{\mathinner
        {\lbrace\,{\mathcode`\|32768\let|\midvert #1}\,\rbrace}}
  \xdef\Set{\protect\expandafter\noexpand\csname Set \endcsname}
  \expandafter\gdef\csname Set \endcsname#1{\left\{%
     \ifx\SavedDoubleVert\relax \let\SavedDoubleVert\|\fi
     \:{\let\|\SetDoubleVert
     \mathcode`\|32768\let|\SetVert
     #1}\:\right\}}
}
\def\midvert{\egroup\mid\bgroup}
\def\SetVert{\@ifnextchar|{\|\@gobble}
    {\egroup\;\mid@vertical\;\bgroup}}
\def\SetDoubleVert{\egroup\;\mid@dblvertical\;\bgroup}

\begingroup
 \edef\@tempa{\meaning\middle}
 \edef\@tempb{\string\middle}
\expandafter \endgroup \ifx\@tempa\@tempb
 \def\mid@vertical{\middle|}
 \def\mid@dblvertical{\middle\SavedDoubleVert}
\else
 \def\mid@vertical{\mskip1mu\vrule\mskip1mu}
 \def\mid@dblvertical{\mskip1mu\vrule\mskip2.5mu\vrule\mskip1mu}
\fi

\renewcommand{\vec}[1]{\boldsymbol{#1}}

\newcommand{\up}{+}
\newcommand{\down}{-}

\newcommand{\norm}[1]{\left\| #1 \right \|}
\newcommand{\mps}[3]{#1^{[#2], #3_{#2}}}
\newcommand{\mpst}[4]{#1^{[#2], #3_{#2}}_{#4}}

\DeclareMathOperator{\tp}{\otimes}
\DeclareMathOperator{\trace}{Tr}

\newcommand{\tensorp}{\otimes}
\newcommand{\id}{\mathds{1}}

\begin{document}
\title{Simulating local measurements on a quantum many body system with
  stochastic matrix product states}
\author{S{\o}ren Gammelmark}
\author{Klaus M{\o}lmer}
\affiliation{Lundbeck Foundation Theoretical Center for
  Quantum System Research, Department of Physics and Astronomy,
  University of Aarhus, DK 8000 Aarhus C, Denmark}

\begin{abstract}
  We demonstrate how to simulate both discrete and continuous
  stochastic evolution of a quantum many body system subject to
  measurements using matrix product states. A particular, but
  generally applicable, measurement model is analyzed and a simple
  representation in terms of matrix product operators is found. The
  technique is exemplified by numerical simulations of the
  anti-ferromagnetic Heisenberg spin-chain model subject to various
  instances of the measurement model. In particular we focus on local
  measurements with small support and non-local measurements which
  induces long range correlations.
\end{abstract}


\maketitle

\section{Introduction}
With the experimental realization of ultra-cold atoms and optical
lattice-systems experimentalists begin to probe various many-body
models accurately, leading to theoretical and accurate experimental
studies of highly non-trivial physical ideas, such as topological
states of matter, frustrated systems and phase transition
phenomena. In these systems experimentalists are not only measuring
macroscopic observables, but are also beginning to measure local
observables \cite{wurtz_experimental_2009, bakr_quantum_2009,
  mekhov_quantum_nondemolition_2009} and correlation-functions
\cite{rom_free_2006}.

Understanding quantum many-body systems is a challenge due to the
large dimensionality of the many-body Hilbert space and the
associated complexity of the states. Not only can it be difficult to
describe a generic many-body state, but finding the physically
relevant states (e.g. diagonalizing the system Hamiltonian), and
quantifying their physical properties can be extremely difficult.\par

Recent progress in the understanding of entanglement and the
complexity of quantum many body systems have led to several simulation
techniques for strongly correlated systems with local interactions
based on matrix product states (MPS) for a one-dimensional lattice
system, projected entangled pair states (PEPS) and variants hereof in
higher dimensions \cite{shi_classical_2006, vidal_class_2008,
  verstraete_density_2004, vidal_efficient_2004,
  verstraete_matrix_2004, verstraete_renormalization_2004,
  daley_time-dependent_2004}, which enable us to calculate ground
states, perform time-evolution of states and calculate
expectation-values of many interesting physical operators
accurately and efficiently.

In this paper we will use the matrix product state techniques to
simulate local and non-local measurements on quantum many-body
systems. Measurements lead to interesting conditioned dynamics and
provide alternative routes to entanglement generation
\cite{matsukevich_bell_2008,srensen_probabilistic_2003} and to quantum
computing \cite{raussendorf_one-way_2001}. Quantifying interaction and
measurement-induced dynamics occurring on the same time-scale is,
however, highly non-trivial.

In section \ref{sec:MPS} we review the formalism and highlight the
basic features of matrix product states. In section
\ref{sec:quantum_theory_of_measurements} we will briefly review the
quantum theory of measurements and introduce the measurement model
studied in this paper. In section \ref{sec:stocMPS} we describe how to
simulate both discrete and continuous measurements on
matrix product states, and in section \ref{sec:numerical} we show some
example simulations. Finally we conclude and discuss further work in
section \ref{sec:conclusion}.

\section{Matrix product states and operators}
\label{sec:MPS}

\begin{figure}
  \centering
  \includegraphics[width=\columnwidth]{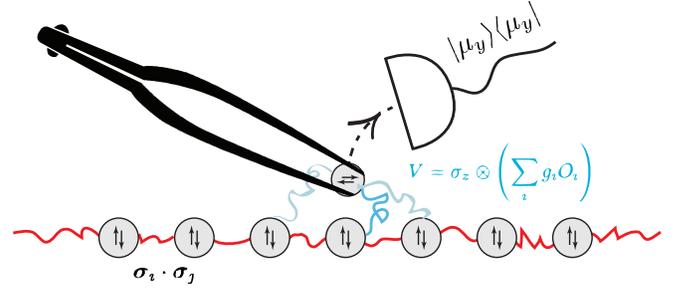}
  \caption{(Color online) Schematic of the measurement setup. A two-level
    ancilla interacts for a time $\tau$ with a specific site or a
    collection of sites in the lattice, after which a projective
    measurement is performed on the ancilla.}
  \label{fig:MeasurementProtocol}
\end{figure}

We wish to study the effect of measurements on many-body systems but,
as mentioned in the introduction, solving many-body problems can be
extremely difficult, and with the added complication of non-equilibrium
stochastic behavior the problem is not getting any easier. We will
describe how to simulate a measurement scheme in terms of \emph{matrix
  product states} (MPS) \cite{vidal_efficient_2004,
  daley_time-dependent_2004} or \emph{projected entangled pair states}
\cite{verstraete_renormalization_2004} (PEPS) (the methods presented here can
be extended very easily from the latter to the former). It will turn
out that local measurements and certain global measurements have a
natural description in terms of matrix product states, and since
measurements are intimately connected to entanglement, this
description will also highlight aspects of the entanglement properties
of matrix product states.\par

First we will briefly review the basics of matrix product states. Any
use of this technique begins with a factorization of the full
many-body Hilbert space into elementary constituents of dimension
$d$. This factorization is done such that the states of interest only
contain a limited amount of entanglement between its subsystems (in
the sense of small Schmidt-number between any bipartite cut). For
spins on a periodic one-dimensional lattice of length $L$ it is
natural to factor the Hilbert-space into a product of single
sites. The MPS ansatz is a parametrization of the expansion
coefficients in the product basis
\begin{align}
  \ket\psi = \sum_{\sigma_1\ldots\sigma_L} \trace\left(
    \mps{T}{1}{\sigma} \mps{T}{2}{\sigma}\ldots
    \mps{T}{L}{\sigma}\right) \bigotimes_{i=1}^L
  \ket{\sigma_i}, \label{eq:mps}
\end{align}
where $\mps{T}{i}{\sigma}$ are $d$ $D\times D$ matrices (or a $d\times
D\times D$ tensor), where $D$ is called the virtual dimension and
$\ket{\sigma_i}$ denote the $d$ single-site basis states. The
trace operation in (\ref{eq:mps}) is carried out over the virtual
dimensions and results in coefficients of the different product
states---the larger the virtual dimension $D$, the more entanglement is
supported by the ansatz.\par

If one wishes to consider a system with open boundary conditions one
can take the first and final matrices $T^{[1]}$ and $T^{[L]}$ to have
dimensions $1\times D$ and $D\times 1$ respectively. The essential
feature of this parametrization is that any Schmidt-decomposition of
the above state will contain at most $D$ terms, thus limiting the
entanglement entropy between any bipartite splitting in a systematic
way. Reciprocally, it can also be shown that any state where all
bipartite splittings contain at most $D$ terms are of this form
\cite{vidal_efficient_2004}.\par

Single site reduced density-matrices for states of this
form can be calculated in a clever way \cite{verstraete_matrix_2008} using
only $O(L d^2 D^3)$ operations. Similarly, few-site density-matrices
can be calculated efficiently. Thus, expectation values of local and
correlation-observables can be calculated efficiently for matrix
product states.

Using variational methods it is possible not only to find good
approximations for ground-states of nearest-neighbor Hamiltonians, but
also Greens-functions, thermal states and time-evolution of these
states, etc. \cite{verstraete_matrix_2008,vidal_efficient_2004}. All these
variational methods essentially rely on the matrix element
$\braket{\chi|A|\psi}$ between any two matrix-product states and an
operator $A$ to be calculated efficiently, which is often possible for
physically relevant operators, like tensor-products of local
operators.

A special set of efficiently contractible operators are the
\emph{matrix product operators} (MPO), which are constructed in the
same way as matrix product states. An MPO is parametrized as
\begin{align}
  A = \sum_{j_1,\ldots,j_L} \trace\left(M^{[1],j_1} \ldots
    M^{[L],j_L}\right) \bigotimes_{i=1}^L X_{j_i},
\end{align}
where $M^{[i],j_i}$ are $D\times D$ matrices and the $N = d^2$
operators $X_j$ constitute a basis for the operators on a single
site. If we choose the matrix-basis as the $X_j$
(i.e. $\ket{\sigma}\bra{\sigma'}$), the matrices $M^{[i],j_i}$ can be
thought of as matrices of operators.

If $A$ is of this form, applying $A$ to $\ket\psi$ in (\ref{eq:mps})
amounts to the update rule
$\mpst{T'}{j}{\sigma}{(\alpha\alpha'),(\beta\beta')} =
\sum_{\sigma_j',i_j} \mpst{T}{j}{\sigma'}{\alpha\beta}
M^{[j],i_j}_{\alpha'\beta'} \braket{\sigma_j|X_{i_j}|\sigma_j'}$,
where $(\alpha\alpha')$ is to be understood as a combined index as in
the Kronecker matrix-product $C_{(ik),(jl)} = A_{ij} B_{kl}$. The
virtual dimension of $A\ket\psi$ will then be the product of the
virtual dimensions of $A$ and $\ket\psi$. In practice however, it is
not desirable and often not necessary to increase the virtual
dimension as one can use a variational method
\cite{verstraete_density_2004} to find a matrix product state
$\ket\chi$ of a given virtual dimension that minimizes $\norm{\ket\chi
  - A\ket\psi}^2$. In, e.g., time-evolution, where $A$ is an
approximation to the unitary time-evolution operator, one is
interested in keeping a fixed virtual dimension, and the error
associated with this truncation is usually negligible for short
times. As shown in \cite{hartmann_density_2009} it is also possible to
perform the time evolution in the Heisenberg picture using matrix
product operators.

The above techniques can be generalized in terms of \emph{projected
  entangled pair states} (PEPS)
\cite{verstraete_renormalization_2004}, where the state is written in
terms of a general tensor-network \cite{cirac_renormalization_2009}
instead of just nearest neighbor contractions on a one-dimensional
lattice. For these kinds of states it is possible to formulate
variational methods in much the same way as for MPS, but the numerical
stability and time-complexity of the methods can depend drastically on
the topology of the graph. For an open one-dimensional lattice with
nearest-neighbor interactions calculating ground states and performing
time-evolution scales as $O(L D^3)$, whereas, for a one-dimensional
lattice with periodic boundary-conditions, the calculations scales as
$O(L D^5)$ due to the loop in the factorization-graph
\cite{markov_simulating_2008}.

The topology of the factorization-graph is usually determined by the
interactions in the system, and may be a chain, a loop, a grid, a tree
or some variation of these depending on the system. Since non-local
measurements can induce entanglement directly between the measured
subsystems due to the obtained information, the topology can also
depend on the type of measurements being performed on the system.

\section{The quantum theory of measurements}
\label{sec:quantum_theory_of_measurements}

If we wish to simulate the stochastic evolution of a quantum system
subject to measurements we can use several, closely related,
techniques depending on the type of measurement being performed.

When a quantum system is being measured, the quantum state $\ket\psi$
changes because the measurement alters the observers knowledge of the
system. Formally this back-action is described by the action of a set
of operators $\{\Omega_\mu\}$, where the state conditioned on the
measurement outcome $\mu$ is $\Omega_\mu \ket\psi /
\sqrt{\braket{\psi|\Omega_\mu^\dagger \Omega_\mu|\psi}}$. In general,
such a set of operators is a valid measurement scheme if $\sum_\mu
\Omega_\mu^\dagger \Omega_\mu = \id$. It can be shown
\cite{nielsen_quantum_2000} that the action of any such set of
operators can be implemented by coupling the measured system unitarily
to an auxiliary system and performing a projective measurement on the
auxiliary system. In particular, consider a two-level ancilla
interacting with the probed system. If $\sigma_z$ acts on the ancilla
space and $A$ is an observable of the probed system with spectral
resolution $A = \sum_a a P_a$ ($P_a$ is the projector onto the
eigenspace associated with the eigenvalue $a$), then an interaction on
the form
\begin{align}
  V = g \sigma_z A, \label{eq:MeasurementInteraction}
\end{align}
applied for a time $\tau$ will result in a time-evolution described by
\begin{align}
  U = \exp(-i \tau V) = \sum_a \mathcal{D}(2\phi a, \vec z) \tensorp
  P_a \label{eq:MeasurementUnitary}
\end{align}
where $\mathcal{D}$ is the spin $1/2$ rotation-operator on the ancilla-spin
system and $\phi = g\tau$. In words, depending on the value of the system
observable $A$, the effective spin ancilla will be rotated an angle
$2\phi a$ around the $z$-axis. If we initialize the ancilla spin in
state $\ket{\up_x}$ pointing in the $x$-direction and subsequently
measure the $y$-component, we thus gain partial information about the
operator $A$. Performing this measurement with the outcome $\mu = \pm
1$ for the $\sigma_y$-readout, corresponds to the action of the
measurement operators
\begin{align}
  \Omega_\mu &= (\bra{\mu_y} \tp \id) U (\ket{+_x}\tp\id) =
  \frac{1}{2}\left( e^{-i\phi A } + i \mu e^{i\phi A}
  \right) \label{eq:MeasurementOperator}
\end{align}
up to an arbitrary phase. Note that since we are coupling the system
to a qubit-ancilla we can maximally gain one classical bit of
information from the system for each measurement. Controlling $\phi$
by varying the interaction time $\tau$ or strength $g$ we can vary the
strength of the measurement. For $\phi = 0$ no measurement is
performed. Continuous measurements with infinitesimal changes of
$\ket\psi$ correspond to the limit of frequent applications of weak
measurements, $\phi \to 0$, while for $\phi \sim 1$ the measurement
back-action becomes large.

We can use this formalism to propagate a system being measured with
some rate $\kappa$ by simply evolving the state $\ket\psi$ unitarily
with system Hamiltonian $H$, for a time $\delta t = 1/\kappa$, and
then apply the measurement operator $\Omega_\mu$: $\ket\psi
\to\Omega_\mu\ket\psi / \sqrt{p(\mu)}$, where we assume the measurement
is much faster than $\delta t$ and $\mu$ is chosen randomly according
to the distribution $p(\mu) = \braket{\psi|\Omega_\mu^\dagger
  \Omega_\mu|\psi}$. This is then repeated until the desired final time.\par

If the system is continuously monitored, we model the measurement
process as a limit of the above procedure in the following sense: When
all the $\Omega_\mu$ are infinitesimally close to the identity, and
the measurements are performed with a large rate compared to the
system dynamics, the accumulated effect of measurements in any given
time-interval (small on the timescale of unitary evolution) amounts to
$N_{\mu}$ applications of $\Omega_\mu$ where $N_\mu$ are binomially
distributed random numbers. Performing the limiting procedure results
in a stochastic differential equation.

To be precise, if $\phi \ll 1$ and $1/\kappa \ll t_{sys}$ then the
accumulated effect of $N = \kappa \Delta T$ measurements in a time
$\Delta T$ is given by
\begin{align}
  \begin{split}
    \Omega(N_+, N_-) \propto \id + \phi A (N_+ - N_-) \\
    + \frac{1}{2} \phi^2 A^2((N_+ - N_-)^2 - 2N).
  \end{split}
\end{align}
In the limit $N \gg 1$ we can apply the central-limit theorem to
obtain $N_+ - N_- \approx 2 \kappa \Delta T \phi \braket{A} +
\sqrt{\kappa} \Delta W$, where $\Delta W$ is a normally distributed
stochastic variable with zero mean and variance $\Delta T$. The
accumulated effect of the measurements can then be written as
\begin{align}
  \begin{split}
    \Omega(N_+, N_-) \propto \id + 2 \Delta T \phi^2\kappa \braket{A}
    A-
    \frac{1}{2} \Delta T \phi^2\kappa A^2 \\
    + \Delta W \sqrt{\phi^2 \kappa } A,
  \end{split}, \label{eq:AccumulatedOperator}
\end{align}
Including the Hamiltonian evolution of the system and
state-normalization, then in the limit $\Delta T \to dt$ we get the
stochastic differential equation
\begin{align}
  \begin{split}
    d\ket\psi = -i H dt\ket\psi + dt \frac{\Gamma}{2}\left( 2\braket{A}A -
      A^2 -
      \braket{A}^2\right)\ket\psi \\
    + \sqrt{\Gamma} (A - \braket{A}) dW \ket\psi,
  \end{split} \label{eq:StochasticContinousMeasurementEquation}
\end{align}
where $\Gamma = \phi^2\kappa$ is the measurement strength.

\section{Stochastic propagation of matrix product states}
\label{sec:stocMPS}

If we consider measurements on a system described by an MPS
then we proceed as described in section
\ref{sec:quantum_theory_of_measurements}: Propagate $\ket\psi$
unitarily for a time $\delta t$ and then apply $\Omega_\mu$ with
probability $p(\mu) = \braket{\psi|\Omega_\mu^\dagger \Omega_\mu|\psi}$. If
$\Omega_\mu$ is efficiently contractible we can use the same variational
principle as used for time-evolution, to find the matrix-product state
of a given virtual dimension which best approximate $\Omega_\mu \ket\psi$.

In particular, if $\Omega_\mu$ can be written as a matrix-product
operator the matrix-element $\braket{\chi|\Omega_\mu|\psi}$ for any MPS
ansatz $\ket\chi$ can be calculated. The probability $p(\mu) =
\braket{\psi|\Omega_\mu^\dagger \Omega_\mu|\psi}$ can also be calculated
efficiently since $\Omega_\mu^\dagger \Omega_\mu$ is also a matrix-product
operator, although possibly of a higher virtual dimension. In
particular if the system is measured at a \emph{single} site,
$\Omega_\mu$ and $\Omega_\mu^\dagger \Omega_\mu$ are just tensor products of
a number of identities and a single-site operator. In this case the
application of $\Omega_\mu$ to the MPS is of course trivial just as the
application of any local operator $V$ is trivial for an MPS: Simply
update $\mps{T'}{i}{\sigma}_{\alpha\beta} =
\sum_{\sigma_i'}\mps{T}{i}{\sigma'}_{\alpha\beta}
\braket{\sigma_i|V|\sigma_i'}$, where $V$ acts on site $i$.\par

We thus find that many-body systems with local interactions combined
with single-site measurements are easy to simulate using matrix
product states in one dimension or PEPS in higher dimensions. If,
however, the measurement extends across multiple sites, e.g. if our
probe is not absolutely confined to a single site, then we need to
decompose $\Omega_\mu$ into an MPO and use the variational methods in
order to avoid growth of the virtual dimension of the state.

We will now consider the measurement discussed in section
\ref{sec:quantum_theory_of_measurements} described by
(\ref{eq:MeasurementInteraction}), (\ref{eq:MeasurementUnitary}) and
(\ref{eq:MeasurementOperator}), but with $A$ a sum of local operators,
i.e. $A = \sum_i g_j O_j$. This can arise either as a combined
simultaneous interaction, but also as result of a sequential
interaction of an ancilla with multiple sites. As in equation
(\ref{eq:MeasurementOperator}) the measurement operators for this
measurement is given by
\begin{align}
  \Omega_\mu  &= \frac{1}{2}\left( e^{-i\phi\sum_j g_j O_j } + i \mu
    e^{i\phi \sum_j g_j O_j} \right) \label{eq:OmegaMgeneral} \\
  &= \frac{1}{2}\left( \bigotimes_j e^{-i\phi g_j O_j} + i \mu
    \bigotimes_j e^{i\phi g_j O_j} \right)
\end{align}
We see that applying this $\Omega_\mu $ to a general MPS results in a
superposition of two MPS each a copy of the original state but
multiplied with the product-operators $\tensorp_j \exp(\pm i\phi g_j
O_j)$. One should think that such a superposition state would not be
well represented by an MPS, but if we increase the virtual dimension
of the state from $D$ to $2D$ we can in fact represent the state exactly
as well as any superposition of two MPSs \cite{mcculloch_density-matrix_2007}.\par

Indeed, $\Omega_\mu$ can be written as an MPO with tensors
\begin{align}
  \begin{split}
    M^{[1]} &=
    \begin{pmatrix}
      \frac{1}{2} \exp(-i\phi g_1 O_1) & \frac{1}{2} i \mu \exp(i \phi g_1 O_1)
    \end{pmatrix} \\
    M^{[i]} &=
    \begin{pmatrix}
      \exp(-i\phi g_i O_i) & 0 \\
      0 & \exp(i \phi g_i O_i)
    \end{pmatrix} \\
    M^{[L]} &=
    \begin{pmatrix}
      \exp(-i\phi g_L O_L) \\
      \exp(i \phi g_L O_L)
    \end{pmatrix},
  \end{split} \label{eq:OmegaMPO}
\end{align}
i.e., of virtual dimension 2. Note that it is the use of a
2-dimensional ancilla, that results in virtual dimension 2. Had we
used an $n$-dimensional ancilla, generically we would need a virtual
dimension of $n$ for the MPO. For the calculation of the branching
probabilities $p(\mu)$, we note that also
$\Omega_\mu^\dagger\Omega_\mu$ can be represented by an MPO of virtual
dimension 3, since $\Omega_\mu^\dagger \Omega_\mu = \frac{1}{2}(\id +
2\mu i(\exp(2i\phi A) - \exp(-2i\phi A)))$.

As mentioned above, the variational method can now be applied to find
a matrix-product state with a specified virtual dimension which is
closest to $\Omega_\mu \ket\psi$. The truncation-error for the
posterior state will naturally depend not only on how close to the
identity $\Omega_\mu$ is, but also on the state being measured. As an
example, an initially uncorrelated spin-chain where all the spins
point in the $x$-direction, will get long-range correlations if we,
e.g., measure the $z$-component of the total spin to 0. If however, a
measurement of the $x$-component is performed, the spin-chain will
remain in a product state.

In the above measurement scheme the rate $\kappa$ is the inverse of
the simulation time-step and the measurement strength $\phi$ can be
adjusted. If we wish to model continuous measurements, the measurement
rate and strength can no longer be chosen independently and the
accumulated effect of many measurements for each simulation time-step
has to be taken into account. One approach would be to implement the
limit discussed in section \ref{sec:quantum_theory_of_measurements} by
simply choosing smaller time-steps and scaling $\phi$ accordingly. But
then a larger number of truncations will be performed, so even if one
is able to apply the measurement exactly the regular time-evolution
may lead to an increased error.

If we seek to approximate the continuous measurement regime, the
time-evolution is given by
(\ref{eq:StochasticContinousMeasurementEquation}), which is not only
non-linear in $\ket\psi$ via the terms $\braket{A}A$ and
$\braket{A}^2$, but also contains terms proportional to $A^2$, which
cannot readily be cast into a matrix-product form. In each time-step,
however, we can calculate $\braket{A}$ and pick a random $\Delta W$ to
select the relevant $\Omega$-operator. To apply this operator to
$\ket\psi$ we need to evaluate $\braket{\chi|\Omega|\psi}$ as
described in section \ref{sec:MPS}, but even though $\Omega$ is not on
a clear MPO-form one can just evaluate the terms separately, provided
$A^2$ is not too pathological. This amounts to the stochastic Euler
method which has a global error of $O(\Delta t^{1/2})$
\cite{kloeden_numerical_1995}.

Alternatively, and in general, if the measurement extends over more
than one site, it is of course always possible to combine those
subsystems into a single Hilbert space and then apply the measurement
operator directly on that space. If the measurement has sufficiently
small support, this will not ruin the power of MPS, since it only
treats a small part of the system exactly.

\section{Numerical examples}
\label{sec:numerical}

\begin{figure}[tb]
  \centering \subfigure
  {\includegraphics{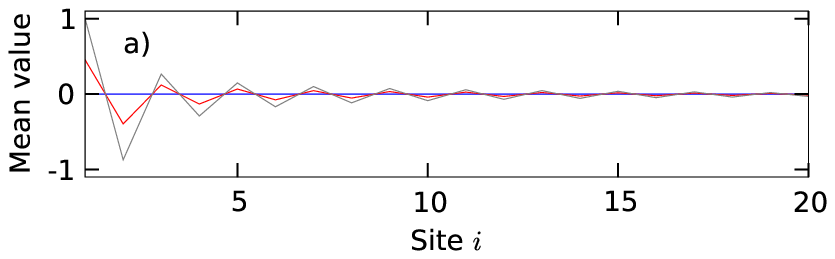} } \\
  \subfigure
  {\includegraphics{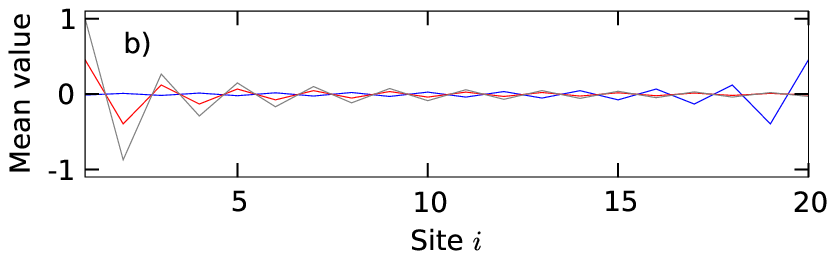} }

  \caption{(Color online) Examples of posterior states after different
    types of measurement on the spin-chain ground state. In both plots
    $\phi = 0.3 \pi/4$ using the measurement operators
    (\ref{eq:MeasurementOperator}). The plotted values are
    $\braket{\sigma_z^i}$ (red) $\braket{\sigma_y^i}$ (blue) and
    $\braket{\sigma_z^1 \sigma_z^i}$ for the ground state (solid
    black) for states subject to $\Omega_+(A = \sigma_z^1)$ in
    \ref{fig:ScenarioExamples}a and $\Omega_+(A =
    \sigma_z^1)\Omega_+(A = \sigma_y^L)$ in
    \ref{fig:ScenarioExamples}b.}
  \label{fig:ScenarioExamples}
\end{figure}

We have applied the above techniques to a spin-1/2 chain with an
anti-ferromagnetic Heisenberg Hamiltonian
\begin{align}
  H = \frac{1}{2} J \sum_{\langle i, j \rangle} \vec \sigma_i \cdot \vec \sigma_j,
\end{align}
where the sum is over nearest neighbors. We first find an MPS for the
ground state, and with this as our initial state we begin to probe the
system with the measurement scheme outlined above. 

\subsection{Local measurements}

The interesting regime in this numerical study is when the
measurement- and Hamiltonian dynamics are of comparable importance,
but let us first imagine a projective measurement of the $z$-component
spin on one of the end-points of the chain. Since the reduced
density-matrix for each site in the anti-ferromagnetic ground-state
are completely mixed (the Hamiltonian is rotational invariant), the
expectation value of the local spin is zero and hence the projective
measurement will have probability $1/2$ for both $\ket{\up_z}$ and
$\ket{\down_z}$. The spin is, however, correlated with its neighbors,
and the spin-expectation-value of the conditioned state
$\braket{\sigma_z^i}_{\mathrm{cond}}$ is directly related to the ground
state correlation function.\par

Rewriting the correlation-function between two arbitrary spins in
terms of the projector onto eigenspaces of the $i$'th spin-$z$,
$P_{\up/\down}^i$, we obtain
\begin{align*}
  \braket{\sigma_z^i \sigma_z^j} =
  \frac{1}{2}\left(\braket{P_\up^i \sigma_z^j} -
    \braket{P_\down^i \sigma_z^j} \right)
  = \braket{P_\up^i \sigma_z^j} = -\braket{P_\down^i \sigma_z^j},
\end{align*}
since the ground state is invariant under $\sigma_z^j \to -\sigma_z^j$
for all $j$. But $\braket{P_{\pm} \sigma_z^j}$ is proportional to the
conditioned expectation value of $\sigma_z^j$.

To get a feel for what a single
measurement (close to the projective case) can do to a spin-chain
ground state see figure \ref{fig:ScenarioExamples}. Figure
\ref{fig:ScenarioExamples}a shows data for the state $\Omega_+\ket\psi$, where
$\Omega_\mu$ is given by (\ref{eq:MeasurementOperator}) with $\phi =
0.3 \pi/4$ and $A = \sigma_z^1$. We see, as noted above, the spins
align (on average) with the $z$-$z$ correlation function and nothing
happens to the $\sigma_y$-average. In addition,
$\braket{\sigma_z^1\sigma_z^i}$ is less modulated than in the initial
ground state due to the partial projection. In figure
\ref{fig:ScenarioExamples}b we show the effects of a joint measurement,
where we couple two ancillas to the system---one to $\sigma_z^1$ and
one to $\sigma_y^L$. Notice how the $z$ and $y$-averages follow their
respective correlation-functions from each end of the spin-chain.

\begin{figure}
  \centering
  \includegraphics{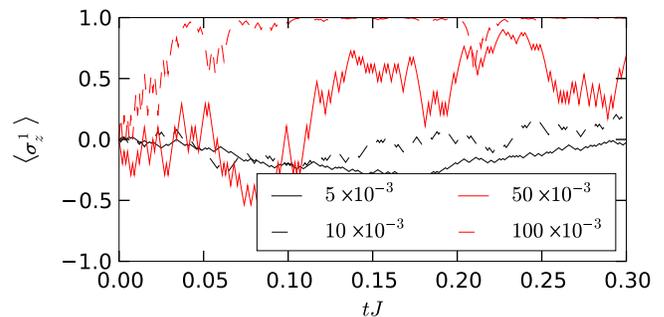}
  \caption{(Color online) Trajectories for single-site measurements with varying
    strengths showing both the quantum zeno- and weak-limit of the
    measurements. In all simulations $\kappa = 100 J$ and the values
    of $\phi$ are shown in the legend.}
  \label{fig:SingleSiteVaryingMstrength}
\end{figure}

If the system- and measurement-dynamics occur on the same timescale
the dynamics becomes far more complex. We have simulated repeated
measurements over time for a spin-chain with 60 spins with varying
measurement strengths $\phi$, as shown in figure
\ref{fig:SingleSiteVaryingMstrength}, where time-series for
$\braket{\sigma_z^1}$ is shown. For weak measurements (black solid
and dashed) the measurements are not strong enough to project the spin
to a $\sigma_z$-eigenstate, since the spin is strongly driven by its
neighbors. For strong measurements (red dashed) the interaction is not
strong enough to drive the measured spin away from its measured value
quickly enough, and we observe a quantum zeno effect on the first
spin; effectively pinning the spin to a random, but definite,
direction. In the intermediate regime (red solid) the measured spin
exhibits oscillations with comparable signatures from both
measurement- and interaction-dynamics.

\begin{figure*}[tb]
  \centering
  \includegraphics{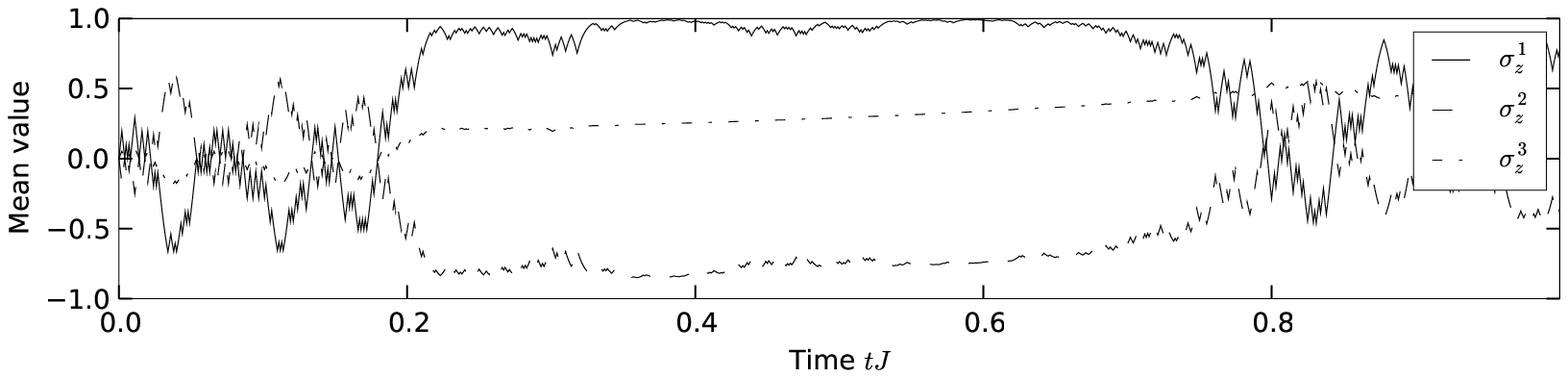}
  \caption{Time-series of $\braket{\sigma_z^1}$ (solid),
    $\braket{\sigma_z^2}$ (dashed) and $\braket{\sigma_z^3}$
    (dash-dotted) for measurements using
    (\ref{eq:MeasurementOperator}) with $A = \sigma_z^1$ for $\phi =
    0.05$ and $\kappa = 100 J$. }
  \label{fig:SingleSiteTimeseries}
\end{figure*}

\begin{figure*}[tb]
  \centering
  \includegraphics{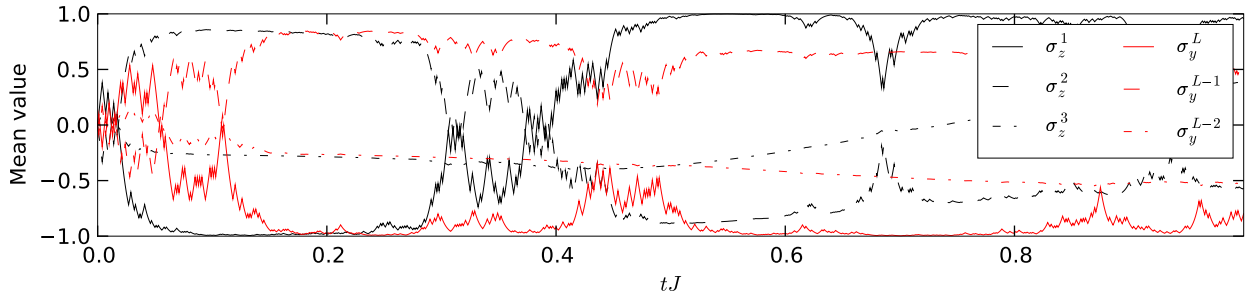}
  \caption{(Color online) Time-series for measurements using
    (\ref{eq:MeasurementOperator}) with both $A = \sigma_z^1$ and $A =
    \sigma_y^L$ for $\phi = 0.05$ and $\kappa = 100 J$.}
  \label{fig:EndpointTimeseries}
\end{figure*}

In figure \ref{fig:SingleSiteTimeseries} the
single-site scheme has been simulated (as in figure
\ref{fig:ScenarioExamples}a) and the expectation-values of $\sigma_z$ for the
first three spins are shown as function of time. As noted above the
measurements induce oscillations in the average-spin values, and in
the example shown, the measurements are strong enough to project the
first spin onto an eigenstate for a long interval of time. Note also
that the second and third spins reflect anti-correlation and
correlation with the first spin respectively, both in time-intervals,
when the spin is almost in a $\sigma_z$-eigenstate, and when
$\braket{\sigma_z^1}$ is close to zero.\par

Notice, that the anti-correlations persist even after the measured
spin has been almost completely projected. This can be understood, if
we consider the weak measurements as a stochastic perturbation: In
order to excite the high-energy non-anti correlated eigenstates of $H$
the stochastic perturbation needs to have significant support at
frequencies of comparable magnitude. Since this particular measurement
is weak it does not provide enough energy to excite these states.

Figure \ref{fig:EndpointTimeseries} shows simulations, where
$\sigma^1_z$ measurements are performed on site 1 and $\sigma_y^L$ at
the end $L = 60$ as in figure \ref{fig:ScenarioExamples}b. Here we also see
the characteristic oscillations of measurements competing with
interactions. In this case the measurements appear to be completely
independent. In the very long time-limit, one might expect to see
temporal correlations arise due to propagation-effects along the
chain.

\subsection{Non-local measurements}

\begin{figure}
  \includegraphics[width=0.9\columnwidth]{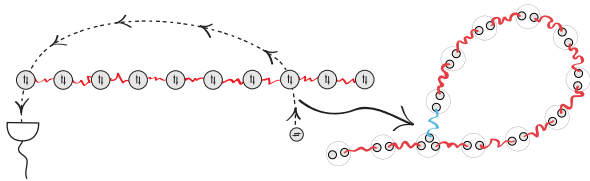}
  \caption{(Color online) To deal with non-local measurements, e.g., by sequential
    system-ancilla interactions (left), in the PEPS-formalism it is
    necessary to add an additional entanglement-bond in order to
    provide an efficient parametrization of the many-body state
    (right).} \label{fig:nonLocal}
\end{figure}

\begin{figure*}
  \centering
  \includegraphics[width=\textwidth]{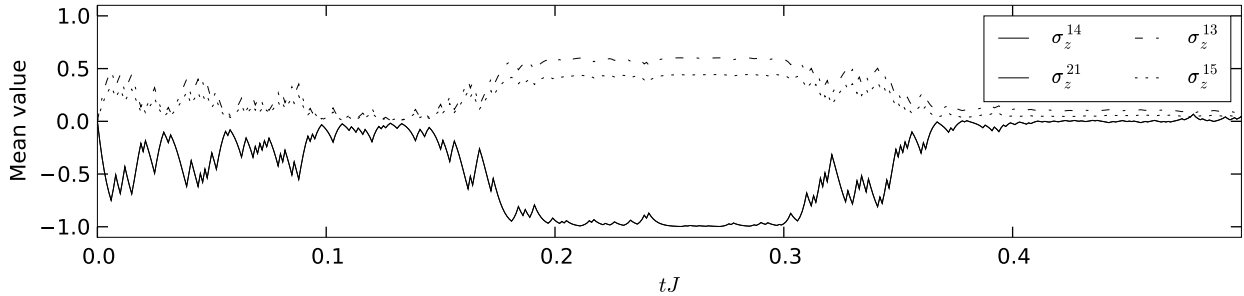}
  \caption{Time-series for a measurement using
    (\ref{eq:MeasurementOperator}) for $A = \sigma_z^{14} +
    \sigma_z^{21}$, $\phi = 0.1$ and $\kappa = 100 J$.}  \label{fig:NonLocalSum}
\end{figure*}

In figure \ref{fig:nonLocal} we consider a process where we measure
two sites far from each other at the same time. This might be
accomplished by an ancilla first interacting with site $i$, then
with site $j$ and then being projectively measured.

This can be simulated in two ways within the theory of MPS and PEPS. One way
is to keep the usual matrix product states and simply apply
(\ref{eq:OmegaMgeneral}) in the form of MPO as in (\ref{eq:OmegaMPO})
where $g_k = 0$ for $k \neq i,j$. The result of such a simulation can
be seen in figure \ref{fig:NonLocalSum}, where we have simulated a
measurement of the sum $\sigma_z^{14} + \sigma_z^{21}$ in a lattice of
30 spins. In the simulation we see many of the same features as in
figures \ref{fig:SingleSiteTimeseries} and
\ref{fig:EndpointTimeseries}, where the spin-expectation values are
projected onto eigenstates of the observable $A$. In this case the
observable $A$ also has eigenstates with eigenvalue 0, and in the
simulations (figure \ref{fig:NonLocalSum}) we see a projection onto
this eigenvalue for $t > 4$. By closer inspection of the simulation
data it was found that in the periods where the measured spins have
average value 0, the two-site density matrix is a slightly asymmetric
statistical mixture of the $\ket{S = 0, m = 0}$ and $\ket{S = 1,
  m = 0}$ state with a purity of about $50\%$.

For sites not too far apart this works well, but truncation-error will
quickly become a problem due to the emerging entanglement between site
$i$ and $j$. If the sites are far apart, however, one might introduce
a PEPS-graph as shown to the right in figure \ref{fig:nonLocal}, where
an extra entanglement bond has been introduced between the two distant
sites. This topology constitutes a new parametrization of our states,
but it is also a good illustration of the physical effects of the
measurement: We are now, effectively dealing with a kind of closed
boundary conditions mediated by measurements. This entails an increase
in computational complexity as mentioned above, but it is still
possible to use many of the same numerical tricks and techniques as
used in open- and closed boundary conditions in usual
MPS-simulations.\par

\section{Conclusion and outlook}
\label{sec:conclusion}
In summary we have shown how to simulate measurements on a quantum
many-body system described by a MPS or PEPS formalism. In particular,
a natural class of measurement-schemes can be represented simply by
matrix product operators and their PEPS-equivalents. We have
illustrated the use of these techniques on the anti-ferromagnetic
Heisenberg spin-chain ground state and investigated the dynamics
resulting from particular weak measurement schemes.

There are a number of natural applications and extensions of this
work. Cold atoms in optical lattices constitute a very attractive
model of many-body dynamics, where both direct optical imaging by a
high aperture lens \cite{bakr_quantum_2009} and by the transmission
properties of an optical cavity enclosing part of the atomic ensemble
\cite{karski_nearest-neighbor_2009, takamizawa_miniature_2006,
  mekhov_quantum_nondemolition_2009,mekhov_quantum_optics_2009,mekhov__ultracold_2009}
is possible. High resolution achieved through non-linear atomic
response \cite{yavuz_nanoscale_2007,gorshkov_coherent_2008} as well as
localized ionization signals, due to the impact of a scanning electron
beam \cite{wurtz_experimental_2009}, may be modelled by our approach.

This will allow studies of the interplay between measurement induced
and interaction induced localization phenomena in such models. In a
future perspective, closed feedback-loops on quantum many body systems
may constitute a promising application to perform more general quantum
state engineering and possibly to control phase transitions in many-body
systems.\par


\begin{thebibliography}{27}
\expandafter\ifx\csname natexlab\endcsname\relax\def\natexlab#1{#1}\fi
\expandafter\ifx\csname bibnamefont\endcsname\relax
  \def\bibnamefont#1{#1}\fi
\expandafter\ifx\csname bibfnamefont\endcsname\relax
  \def\bibfnamefont#1{#1}\fi
\expandafter\ifx\csname citenamefont\endcsname\relax
  \def\citenamefont#1{#1}\fi
\expandafter\ifx\csname url\endcsname\relax
  \def\url#1{\texttt{#1}}\fi
\expandafter\ifx\csname urlprefix\endcsname\relax\def\urlprefix{URL }\fi
\providecommand{\bibinfo}[2]{#2}
\providecommand{\eprint}[2][]{\url{#2}}

\bibitem[{\citenamefont{Wurtz et~al.}(2009)\citenamefont{Wurtz, Langen,
  Gericke, Koglbauer, and Ott}}]{wurtz_experimental_2009}
\bibinfo{author}{\bibfnamefont{P.}~\bibnamefont{Wurtz}},
  \bibinfo{author}{\bibfnamefont{T.}~\bibnamefont{Langen}},
  \bibinfo{author}{\bibfnamefont{T.}~\bibnamefont{Gericke}},
  \bibinfo{author}{\bibfnamefont{A.}~\bibnamefont{Koglbauer}},
  \bibnamefont{and} \bibinfo{author}{\bibfnamefont{H.}~\bibnamefont{Ott}},
  \bibinfo{journal}{Phys. Rev. Lett.} \textbf{\bibinfo{volume}{103}},
  \bibinfo{pages}{080404} (\bibinfo{year}{2009}).

\bibitem[{\citenamefont{Bakr et~al.}(2009)\citenamefont{Bakr, Gillen, Peng,
  Foelling, and Greiner}}]{bakr_quantum_2009}
\bibinfo{author}{\bibfnamefont{W.~S.} \bibnamefont{Bakr}},
  \bibinfo{author}{\bibfnamefont{J.~I.} \bibnamefont{Gillen}},
  \bibinfo{author}{\bibfnamefont{A.}~\bibnamefont{Peng}},
  \bibinfo{author}{\bibfnamefont{S.}~\bibnamefont{Foelling}}, \bibnamefont{and}
  \bibinfo{author}{\bibfnamefont{M.}~\bibnamefont{Greiner}},
  \bibinfo{journal}{arXiv:0908.0174 [cond-mat]}  (\bibinfo{year}{2009}).

\bibitem[{\citenamefont{Mekhov and
  Ritsch}(2009{\natexlab{a}})}]{mekhov_quantum_nondemolition_2009}
\bibinfo{author}{\bibfnamefont{I.~B.} \bibnamefont{Mekhov}} \bibnamefont{and}
  \bibinfo{author}{\bibfnamefont{H.}~\bibnamefont{Ritsch}},
  \bibinfo{journal}{Phys. Rev. Lett.} \textbf{\bibinfo{volume}{102}},
  \bibinfo{pages}{020403} (\bibinfo{year}{2009}{\natexlab{a}}).

\bibitem[{\citenamefont{Rom et~al.}(2006)\citenamefont{Rom, Best, van Oosten,
  Schneider, Folling, Paredes, and Bloch}}]{rom_free_2006}
\bibinfo{author}{\bibfnamefont{T.}~\bibnamefont{Rom}},
  \bibinfo{author}{\bibfnamefont{T.}~\bibnamefont{Best}},
  \bibinfo{author}{\bibfnamefont{D.}~\bibnamefont{van Oosten}},
  \bibinfo{author}{\bibfnamefont{U.}~\bibnamefont{Schneider}},
  \bibinfo{author}{\bibfnamefont{S.}~\bibnamefont{Folling}},
  \bibinfo{author}{\bibfnamefont{B.}~\bibnamefont{Paredes}}, \bibnamefont{and}
  \bibinfo{author}{\bibfnamefont{I.}~\bibnamefont{Bloch}},
  \bibinfo{journal}{Nature} \textbf{\bibinfo{volume}{444}},
  \bibinfo{pages}{733} (\bibinfo{year}{2006}).

\bibitem[{\citenamefont{Shi et~al.}(2006)\citenamefont{Shi, Duan, and
  Vidal}}]{shi_classical_2006}
\bibinfo{author}{\bibfnamefont{Y.~Y.} \bibnamefont{Shi}},
  \bibinfo{author}{\bibfnamefont{L.~M.} \bibnamefont{Duan}}, \bibnamefont{and}
  \bibinfo{author}{\bibfnamefont{G.}~\bibnamefont{Vidal}},
  \bibinfo{journal}{Phys. Rev. A} \textbf{\bibinfo{volume}{74}},
  \bibinfo{pages}{022320} (\bibinfo{year}{2006}).

\bibitem[{\citenamefont{Vidal}(2008)}]{vidal_class_2008}
\bibinfo{author}{\bibfnamefont{G.}~\bibnamefont{Vidal}},
  \bibinfo{journal}{Phys. Rev. Lett.} \textbf{\bibinfo{volume}{101}},
  \bibinfo{pages}{110501} (\bibinfo{year}{2008}).

\bibitem[{\citenamefont{Verstraete
  et~al.}(2004{\natexlab{a}})\citenamefont{Verstraete, Porras, and
  Cirac}}]{verstraete_density_2004}
\bibinfo{author}{\bibfnamefont{F.}~\bibnamefont{Verstraete}},
  \bibinfo{author}{\bibfnamefont{D.}~\bibnamefont{Porras}}, \bibnamefont{and}
  \bibinfo{author}{\bibfnamefont{J.~I.} \bibnamefont{Cirac}},
  \bibinfo{journal}{Phys. Rev. Lett.} \textbf{\bibinfo{volume}{93}},
  \bibinfo{pages}{227205} (\bibinfo{year}{2004}{\natexlab{a}}).

\bibitem[{\citenamefont{Vidal}(2004)}]{vidal_efficient_2004}
\bibinfo{author}{\bibfnamefont{G.}~\bibnamefont{Vidal}},
  \bibinfo{journal}{Phys. Rev. Lett.} \textbf{\bibinfo{volume}{93}},
  \bibinfo{pages}{040502} (\bibinfo{year}{2004}).

\bibitem[{\citenamefont{Verstraete
  et~al.}(2004{\natexlab{b}})\citenamefont{Verstraete, {Garcia-Ripoll}, and
  Cirac}}]{verstraete_matrix_2004}
\bibinfo{author}{\bibfnamefont{F.}~\bibnamefont{Verstraete}},
  \bibinfo{author}{\bibfnamefont{J.~J.} \bibnamefont{{Garcia-Ripoll}}},
  \bibnamefont{and} \bibinfo{author}{\bibfnamefont{J.~I.} \bibnamefont{Cirac}},
  \bibinfo{journal}{Phys. Rev. Lett.} \textbf{\bibinfo{volume}{93}},
  \bibinfo{pages}{207204} (\bibinfo{year}{2004}{\natexlab{b}}).

\bibitem[{\citenamefont{Verstraete and
  Cirac}(2004)}]{verstraete_renormalization_2004}
\bibinfo{author}{\bibfnamefont{F.}~\bibnamefont{Verstraete}} \bibnamefont{and}
  \bibinfo{author}{\bibfnamefont{J.~I.} \bibnamefont{Cirac}},
  \bibinfo{journal}{arXiv:0407066 [cond-mat]}  (\bibinfo{year}{2004}).

\bibitem[{\citenamefont{Daley et~al.}(2004)\citenamefont{Daley, Kollath,
  Schollwock, and Vidal}}]{daley_time-dependent_2004}
\bibinfo{author}{\bibfnamefont{A.~J.} \bibnamefont{Daley}},
  \bibinfo{author}{\bibfnamefont{C.}~\bibnamefont{Kollath}},
  \bibinfo{author}{\bibfnamefont{U.}~\bibnamefont{Schollwock}},
  \bibnamefont{and} \bibinfo{author}{\bibfnamefont{G.}~\bibnamefont{Vidal}},
  \bibinfo{journal}{Journal of Statistical Mechanics: Theory and Experiment}
  \textbf{\bibinfo{volume}{2004}}, \bibinfo{pages}{P04005}
  (\bibinfo{year}{2004}).

\bibitem[{\citenamefont{Matsukevich et~al.}(2008)\citenamefont{Matsukevich,
  Maunz, Moehring, Olmschenk, and Monroe}}]{matsukevich_bell_2008}
\bibinfo{author}{\bibfnamefont{D.~N.} \bibnamefont{Matsukevich}},
  \bibinfo{author}{\bibfnamefont{P.}~\bibnamefont{Maunz}},
  \bibinfo{author}{\bibfnamefont{D.~L.} \bibnamefont{Moehring}},
  \bibinfo{author}{\bibfnamefont{S.}~\bibnamefont{Olmschenk}},
  \bibnamefont{and} \bibinfo{author}{\bibfnamefont{C.}~\bibnamefont{Monroe}},
  \bibinfo{journal}{Phys. Rev. Lett.} \textbf{\bibinfo{volume}{100}},
  \bibinfo{pages}{150404} (\bibinfo{year}{2008}).

\bibitem[{\citenamefont{S{\o}rensen and
  M{\o}lmer}(2003)}]{srensen_probabilistic_2003}
\bibinfo{author}{\bibfnamefont{A.~S.} \bibnamefont{S{\o}rensen}}
  \bibnamefont{and}
  \bibinfo{author}{\bibfnamefont{K.}~\bibnamefont{M{\o}lmer}},
  \bibinfo{journal}{Phys. Rev. Lett.} \textbf{\bibinfo{volume}{90}},
  \bibinfo{pages}{127903} (\bibinfo{year}{2003}).

\bibitem[{\citenamefont{Raussendorf and
  Briegel}(2001)}]{raussendorf_one-way_2001}
\bibinfo{author}{\bibfnamefont{R.}~\bibnamefont{Raussendorf}} \bibnamefont{and}
  \bibinfo{author}{\bibfnamefont{H.~J.} \bibnamefont{Briegel}},
  \bibinfo{journal}{Phys. Rev. Lett.} \textbf{\bibinfo{volume}{86}},
  \bibinfo{pages}{5188} (\bibinfo{year}{2001}).

\bibitem[{\citenamefont{Verstraete et~al.}(2008)\citenamefont{Verstraete, Murg,
  and Cirac}}]{verstraete_matrix_2008}
\bibinfo{author}{\bibfnamefont{F.}~\bibnamefont{Verstraete}},
  \bibinfo{author}{\bibfnamefont{V.}~\bibnamefont{Murg}}, \bibnamefont{and}
  \bibinfo{author}{\bibfnamefont{J.~I.} \bibnamefont{Cirac}},
  \bibinfo{journal}{Adv. Phys.} \textbf{\bibinfo{volume}{57}},
  \bibinfo{pages}{143} (\bibinfo{year}{2008}).

\bibitem[{\citenamefont{Hartmann et~al.}(2009)\citenamefont{Hartmann, Prior,
  Clark, and Plenio}}]{hartmann_density_2009}
\bibinfo{author}{\bibfnamefont{M.~J.} \bibnamefont{Hartmann}},
  \bibinfo{author}{\bibfnamefont{J.}~\bibnamefont{Prior}},
  \bibinfo{author}{\bibfnamefont{S.~R.} \bibnamefont{Clark}}, \bibnamefont{and}
  \bibinfo{author}{\bibfnamefont{M.~B.} \bibnamefont{Plenio}},
  \bibinfo{journal}{Phys. Rev. Lett.} \textbf{\bibinfo{volume}{102}},
  \bibinfo{pages}{057202} (\bibinfo{year}{2009}).

\bibitem[{\citenamefont{Cirac and
  Verstraete}(2009)}]{cirac_renormalization_2009}
\bibinfo{author}{\bibfnamefont{J.~I.} \bibnamefont{Cirac}} \bibnamefont{and}
  \bibinfo{author}{\bibfnamefont{F.}~\bibnamefont{Verstraete}},
  \bibinfo{journal}{arXiv:0910.1130 [cond-mat]}  (\bibinfo{year}{2009}).

\bibitem[{\citenamefont{Markov and Shi}(2008)}]{markov_simulating_2008}
\bibinfo{author}{\bibfnamefont{I.~L.} \bibnamefont{Markov}} \bibnamefont{and}
  \bibinfo{author}{\bibfnamefont{Y.}~\bibnamefont{Shi}},
  \bibinfo{journal}{{SIAM} Journal on Computing} \textbf{\bibinfo{volume}{38}},
  \bibinfo{pages}{963} (\bibinfo{year}{2008}).

\bibitem[{\citenamefont{Nielsen and Chuang}(2000)}]{nielsen_quantum_2000}
\bibinfo{author}{\bibfnamefont{M.~A.} \bibnamefont{Nielsen}} \bibnamefont{and}
  \bibinfo{author}{\bibfnamefont{I.~L.} \bibnamefont{Chuang}},
  \emph{\bibinfo{title}{Quantum computation and quantum information}}
  (\bibinfo{publisher}{Cambridge University Press}, \bibinfo{year}{2000}).

\bibitem[{\citenamefont{McCulloch}(2007)}]{mcculloch_density-matrix_2007}
\bibinfo{author}{\bibfnamefont{I.~P.} \bibnamefont{McCulloch}},
  \bibinfo{journal}{Journal of Statistical Mechanics: Theory and Experiment}
  \textbf{\bibinfo{volume}{2007}}, \bibinfo{pages}{P10014}
  (\bibinfo{year}{2007}).

\bibitem[{\citenamefont{Kloeden and Platen}(1995)}]{kloeden_numerical_1995}
\bibinfo{author}{\bibfnamefont{P.~E.} \bibnamefont{Kloeden}} \bibnamefont{and}
  \bibinfo{author}{\bibfnamefont{E.}~\bibnamefont{Platen}},
  \emph{\bibinfo{title}{Numerical solution of stochastic differential
  equations}} (\bibinfo{publisher}{Springer}, \bibinfo{year}{1995}).

\bibitem[{\citenamefont{Karski et~al.}(2009)\citenamefont{Karski, Forster,
  Choi, Alt, Widera, and Meschede}}]{karski_nearest-neighbor_2009}
\bibinfo{author}{\bibfnamefont{M.}~\bibnamefont{Karski}},
  \bibinfo{author}{\bibfnamefont{L.}~\bibnamefont{Forster}},
  \bibinfo{author}{\bibfnamefont{J.~M.} \bibnamefont{Choi}},
  \bibinfo{author}{\bibfnamefont{W.}~\bibnamefont{Alt}},
  \bibinfo{author}{\bibfnamefont{A.}~\bibnamefont{Widera}}, \bibnamefont{and}
  \bibinfo{author}{\bibfnamefont{D.}~\bibnamefont{Meschede}},
  \bibinfo{journal}{Phys. Rev. Lett.} \textbf{\bibinfo{volume}{102}},
  \bibinfo{pages}{053001} (\bibinfo{year}{2009}).

\bibitem[{\citenamefont{Takamizawa et~al.}(2006)\citenamefont{Takamizawa,
  Steinmetz, Delhuille, H{\"a}nsch, and Reichel}}]{takamizawa_miniature_2006}
\bibinfo{author}{\bibfnamefont{A.}~\bibnamefont{Takamizawa}},
  \bibinfo{author}{\bibfnamefont{T.}~\bibnamefont{Steinmetz}},
  \bibinfo{author}{\bibfnamefont{R.}~\bibnamefont{Delhuille}},
  \bibinfo{author}{\bibfnamefont{T.~W.} \bibnamefont{H{\"a}nsch}},
  \bibnamefont{and} \bibinfo{author}{\bibfnamefont{J.}~\bibnamefont{Reichel}},
  \bibinfo{journal}{Optics Express} \textbf{\bibinfo{volume}{14}},
  \bibinfo{pages}{10976} (\bibinfo{year}{2006}).

\bibitem[{\citenamefont{Mekhov and
  Ritsch}(2009{\natexlab{b}})}]{mekhov_quantum_optics_2009}
\bibinfo{author}{\bibfnamefont{I.~B.} \bibnamefont{Mekhov}} \bibnamefont{and}
  \bibinfo{author}{\bibfnamefont{H.}~\bibnamefont{Ritsch}},
  \bibinfo{journal}{Phys. Rev. A} \textbf{\bibinfo{volume}{80}},
  \bibinfo{pages}{013604} (\bibinfo{year}{2009}{\natexlab{b}}).

\bibitem[{\citenamefont{Mekhov and
  Ritsch}(2009{\natexlab{c}})}]{mekhov__ultracold_2009}
\bibinfo{author}{\bibfnamefont{I.~B.} \bibnamefont{Mekhov}} \bibnamefont{and}
  \bibinfo{author}{\bibfnamefont{H.}~\bibnamefont{Ritsch}}
  (\bibinfo{year}{2009}{\natexlab{c}}), \eprint{arXiv:0911.0389 [quant-ph]}.

\bibitem[{\citenamefont{Yavuz and Proite}(2007)}]{yavuz_nanoscale_2007}
\bibinfo{author}{\bibfnamefont{D.~D.} \bibnamefont{Yavuz}} \bibnamefont{and}
  \bibinfo{author}{\bibfnamefont{N.~A.} \bibnamefont{Proite}},
  \bibinfo{journal}{Phys. Rev. A} \textbf{\bibinfo{volume}{76}},
  \bibinfo{pages}{041802(R)} (\bibinfo{year}{2007}).

\bibitem[{\citenamefont{Gorshkov et~al.}(2008)\citenamefont{Gorshkov, Jiang,
  Greiner, Zoller, and Lukin}}]{gorshkov_coherent_2008}
\bibinfo{author}{\bibfnamefont{A.~V.} \bibnamefont{Gorshkov}},
  \bibinfo{author}{\bibfnamefont{L.}~\bibnamefont{Jiang}},
  \bibinfo{author}{\bibfnamefont{M.}~\bibnamefont{Greiner}},
  \bibinfo{author}{\bibfnamefont{P.}~\bibnamefont{Zoller}}, \bibnamefont{and}
  \bibinfo{author}{\bibfnamefont{M.~D.} \bibnamefont{Lukin}},
  \bibinfo{journal}{Phys. Rev. Lett.} \textbf{\bibinfo{volume}{100}},
  \bibinfo{pages}{093005} (\bibinfo{year}{2008}).

\end{thebibliography}
 
\end{document}